\documentclass[aps,pre,onecolumn,epsf,graphicx]{revtex4}
\def\bacol{\setlength{\arraycolsep}{0pt}}
\def\bec{\begin{center}}
\def\enc{\end{center}}

\def\ben{\begin{equation}}
\def\ba{\begin{array}}
\def\bea{\begin{eqnarray}}

\def\een{\end{equation}}
\def\eea{\end{eqnarray}}
\def\ea{\end{array}}
\def\btab{\begin{table}}
\def\btabu{\begin{tabular}}
\def\etab{\end{table}}
\def\etabu{\end{tabular}}
\def\bit{\begin{itemize}}
\def\eit{\end{itemize}}
\def\bef{\begin{figure}[htb]}
\def\befh{\begin{figure}[!h!]}
\def\enf{\end{figure}}

\def\b1{{\bf 1}}

\def\cos{\hbox{cos}\:}

\def\cosh{\,\hbox{cosh}\;}
\def\sinh{\,\hbox{sinh}\;}

\def\nn{\nonumber}

\usepackage{epsfig,latexsym,subfigure}

\newcommand \bew {\begin{widetext}}
\newcommand \enw {\end{widetext}}

\begin{document}

\title{\bf\noindent Field theoretic calculation of 
the surface tension for a model electrolyte system}

\author{D.S. Dean$^{(1,2)}$ and R.R. Horgan$^{(1)}$}

\affiliation{
(1) DAMTP, CMS, University of Cambridge, Cambridge, CB3 0WA, UK \\
(2) IRSAMC, Laboratoire de Physique Quantique, Universit\'e Paul Sabatier, 118 route de Narbonne, 31062 Toulouse Cedex 04, France\\
E-Mail:dean@irsamc.ups-tlse.fr, rrh@damtp.cam.ac.uk}

\date{16 July 2003}
\begin{abstract}
We carry out the calculation of the surface tension for a model
electrolyte to first order in a cumulant expansion about a free field
theory equivalent to the Debye-H\"uckel approximation. 
In contrast with previous calculations, the surface tension is calculated
directly without recourse to integrating thermodynamic relations.
The system considered is
a monovalent electrolyte with a region at the interface, of width $h$,
from which the ionic species are excluded. In the case where the external 
dielectric constant $\epsilon_0$ is smaller than the electrolyte solution's
dielectric constant $\epsilon$ we show that the calculation at this order
can be fully regularized. In the case where $h$ is taken to be zero the
Onsager-Samaras limiting law for the excess surface tension of dilute
electrolyte solutions is recovered, with corrections coming from a 
non-zero value of $\epsilon_0/\epsilon$.     
\end{abstract}  
\maketitle
\vspace{.2cm}
\pagenumbering{arabic}
\section{Introduction}
The first experiments to measure the surface tension of electrolyte 
solutions show that the excess surface tension, denoted in this paper by $\sigma_e$, due to the presence
of the electrolyte is positive \cite{expsold}. This result has been 
confirmed by more recent experiments \cite{expsnew}. This effect was 
explained by Wagner \cite{wag} who pointed out that when the dielectric
constant of the bulk solvent (here water) $\epsilon$  is greater than
that of the exterior (here air) $\epsilon_0$ then the image charges,
due to the dielectric variation across the surface, repel the solute ions from 
the surface and thus lead to a reduction of the density of ions near the 
surface with respect to the bulk. Applying the Gibbs adsorption
isotherm we then find that $\sigma_e$ must be positive. 
In addition experimental results on systems at weak dilution for solutes
of the same valency are very similar, suggesting a universal limiting
law at weak dilution. Such a universal limiting law was subsequently 
obtained by Onsager and Samaras \cite{onsa}. 

A series of experiments carried out in the 1930s \cite{rj} 
caused a certain controversy
as at very small electrolyte concentrations a negative excess surface 
tension was reported. It seems that these experiments have not been revisited
using modern techniques, or at least have not been reproduced since. If
a negative  $\sigma_e$ is found, then appealing to the Gibbs
adsorption isotherm, there must be some mechanism causing ions to be 
positively adsorbed near the interface. Various authors have discussed
ion-specific effects which could explain such a phenomenon 
\cite{dole,niya,bowini,vlvo,kara}, and also lead to the ion dependent variations
seen in the measurements of $\sigma_e$ at higher concentrations.

The calculation of the surface tension of electrolytes was recently revisited
in a series of papers by Levin \cite{lev} and Levin and Flores-Mena \cite{levfm}. 
Because of the thermodynamic equivalence of ensembles, an exact calculation of the 
surface tension should give the same result independent of the ensemble chosen since
the thermodynamic identities from which the surface tension is calculated are exact. 
However, Levin points out that calculations of the surface tension invariably 
rely on approximation schemes, notably the Debye-H\"uckel approximation,
and that a given approximation scheme will generally yield different results for different 
choices of thermodynamic ensemble. For example, Levin applies a canonical approach whereas
the original Onsager-Samaras result was obtained using the grand canonical ensemble. 
In the approach of Levin $\sigma_e$ is given by the excess Helmholtz free energy
due to the presence of an interface. This free energy excess is obtained by calculating the 
internal energy due to the presence of the interface and then integrating it via the G\"untelberg  
charging process to obtain the free energy. In the limit of weak electrolytes the Onsager-Samaras 
limiting law is  recovered thus, as Levin remarks, suggesting that the Onsager-Samaras
limiting law is indeed exact.

In this paper we calculate $\sigma_e$ in the grand-canonical
ensemble by directly calculating the excess grand potential due to the presence
of an interface. In this way we avoid the integration of differential 
thermodynamic identities such as the Gibbs adsorption isotherm or the 
G\"untelberg charging process, and so provide another route for doing the calculation. 
In addition, we develop a controlled perturbation theory based on a cumulant expansion,
similar to that used for bulk electrolytes by Netz and Orland \cite{neor};
this is a perturbation expansion in the coupling constant $g =l_B/l_D$,
where $l_D$ is the Debye length and $l_B$ the Bjerrum length. We show
that the Onsager-Samaras limiting law is the first term in this cumulant 
expansion, showing that it is indeed exact to this order.
The limiting laws obtained in the literature are given in the limit where
$\epsilon_0/\epsilon \to 0$, which is clearly a good approximation for aqueous
solutions in air where  $\epsilon_0/\epsilon \approx 1/80$. In this paper we generalize
the Onsager-Samaras result and give the corresponding limiting law in the case where 
$\epsilon_0/\epsilon> 0$.  

Our approach is also applied to a modified model of the 
interface where there is surface-exclusion layer for the ions of thickness $h$: 
a region at the surface from which the hydrated ions are forbidden \cite{onsa,lev,rand}. 
Highly accurate numerical integration is used to investigate
the importance of the effect of this exclusion layer on the value of $\sigma_e$.

The techniques used in this paper are based on the field theoretic Sine-Gordon
representation of the grand partition function first introduced in this 
context in \cite{ft1}. The perturbation theory about
the free field or Debye-H\"uckel theory is carried out using a functional
path integral technique introduced recently by the authors 
\cite{deho,dehown,dehocon}, which lends itself to the geometry of planar 
systems and gives a powerful alternative method for the calculation of the 
functional determinants involved. 

We conclude with a discussion of our results and the possible advantages of
our approach for calculating $\sigma_e$ in more complex models where,
for example, a surface charge exists due to a thermodynamic adsorption
process for one of the ionic species at the surface or due to a difference
in the hydrated radii between the cations and anions.

\section{The Model}
\begin{figure}
\epsfxsize=0.7\hsize
\epsfbox{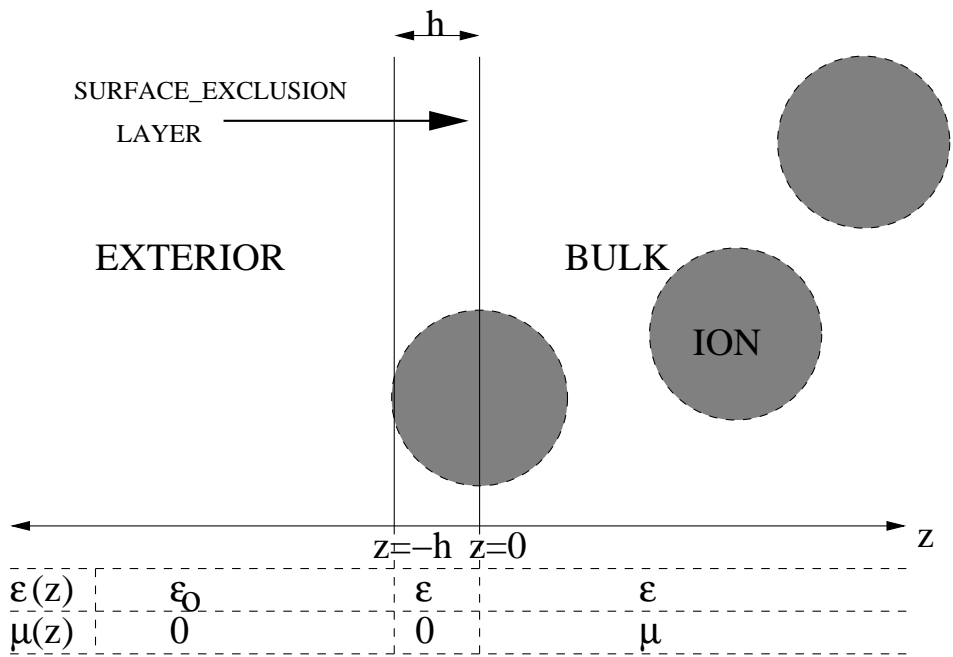}
\caption{Schematic image of the exterior bulk interface for the 
model considered here. The values of the local dielectric constants and 
fugacities as a function of the distance from the dividing surface are shown. The charges of the ions are taken to be at their centers which  are excluded from
the surface-exclusion layer of width $h$.}
\label{film}
\end{figure}

We consider a model consisting of a semi-infinite electrolyte bulk 
with monovalent salt in contact with a semi-infinite exterior,
see Fig. (\ref{film}). The bulk solvent's dielectric constant is denoted by $\epsilon$ and the
exterior dielectric constant is denoted by $\epsilon_0$. There is a
region of width $h$ between the exterior and bulk which is filled
with the bulk solvent but from where the salt ions are excluded, this is
a standard surface-exclusion  layer and was first introduced by Randles \cite{rand} in the 
context of electrolyte surface tensions. The  width $h$ of the surface-exclusion layer 
is the order of a hydrated ion radius. For simplicity both hydrated anions and cations 
are taken to be of the same size and are hence both excluded from this region and so there
is no surface charging process. However, the model and approach can be generalized to
ions of different radii which will lead to different ion-specific surface layer widths
and so allowing a charging mechanism \cite{dehown}. In the bulk solution the fugacity of the cations and
anions is equal and denoted by $\mu$. The system up can be summarized in terms
of a spatially dependent dielectric constant $\epsilon(z)$ and spatially 
dependent fugacity $\mu(z)$ which are defined as follows
\begin{eqnarray}
\epsilon(z) &=& \epsilon_0 \ \ z<-h \nonumber \\
\epsilon(z) &=& \epsilon \ \ \   z >-h 
\end{eqnarray}
and 
\begin{eqnarray}
\mu(z) &=& 0\ \ z< 0 \nonumber \\
\mu(z) &=& \mu \ \ z > 0 
\end{eqnarray}

In the grand canonical ensemble the grand partition function
for the system is given by the functional integral over the Wick rotated 
electrostatic potential $\phi$  
\begin{equation}
\Xi = \int d[\phi] \exp\left( S[\phi] \right)
\end{equation}
with the action $S$ given by
\begin{equation}
S[\phi] = -{\beta\over 2} \int d{\bf x} \ \epsilon({\bf x})
(\nabla \phi)^2 + 2 \int d{\bf x} \ \mu ({\bf x})
\cos\left(e\beta \phi\right)
\end{equation}
where $e$ is the electron charge and $\beta$ is the inverse 
temperature. We take the area of the system across the interface
to be $A$ and the length (in the $z$ direction) of the exterior to be
$L$ and of the bulk to be $L'$. If one considers just the exterior
system without any interface, its  grand partition function is
given by
\begin{equation}
\Xi_{E} = \int d[\phi] \exp\left( S_E[\phi] \right)
\end{equation}
with $S_E$ given by
\begin{equation}
S_E[\phi]= -{\beta\over 2} \int d{\bf x} \ \epsilon_0
(\nabla \phi)^2
\end{equation}
the integration being over the region $L'\times A$. For a pure bulk system with
no interface the grand partition function is given by
\begin{equation}
\Xi_{B} = \int d[\phi] \exp\left( S_B[\phi] \right)
\end{equation}
where $S_B$ is given by
\begin{equation}
S_B[\phi] = -{\beta\over 2} \int d{\bf x} \ \epsilon
(\nabla \phi)^2 + 2 \int d{\bf x} \ \mu
\cos\left(e\beta \phi\right)
\end{equation}
the integration being over the region $L \times A$. The surface tension
is then given by the difference in the grand potential of the system
with the interface and that of the sum of the two individual 
(exterior and bulk) systems divided by the total area, {\em i.e.}
\begin{equation}
\sigma = {1\over A}( J(L',L) - J^{(B)}(L) - J^{(E)}(L'))
\label{dst}
\end{equation}
where $J^{(E)}= -\ln(\Xi_E)/\beta $ and $J^{(B)}= -\ln(\Xi_B)/\beta $ 
denote the grand potentials for
a bulk system of electrolyte and exterior system of the same volumes
$[L\times A]$ and $[L'\times A]$ respectively,
but with no interfaces. The definition of Eq. (\ref{dst}) for the surface
tension is of course also in agreement with various other methods for
calculation. For example it is the same as that obtained from the
Gibbs adsorption equation as originally used by Onsager and Samaras.
The  expression Eq. (\ref{dst}) for the surface tension 
can also be obtained from the formula
\begin{equation}
2\sigma = -\int_0^\infty P_d(L) dL
\end{equation}
where $P_d(L)$ is the disjoining pressure for a film of external medium of thickness $L$
surrounded by bulk electrolyte \cite{is}. This system consists of two bulk surfaces a distance 
$L$ apart, and so twice the surface tension is given by the work needed to create an infinitely 
thick film: $L \to \infty$. As mentioned in the introduction, the approach here is different 
from  previous techniques since here the grand potential difference corresponding to the surface 
tension is calculated directly. 

The excess surface tension $\sigma_e$ for a system with bulk electrolyte
concentration $\rho$ is defined by
\begin{equation}
\sigma_{e}(\rho) = \sigma(\rho) - \sigma(0)
\end{equation} 
where $\sigma(0)$ is the surface tension of the system with no added electrolyte. 
This definition means that $\sigma_e$ is free of the ultraviolet or short-distance
divergences found in calculations of the surface tension between
two media of differing dielectric constants \cite{is,mani}.

In electrostatic problems where the chemical potential and dielectric
constants depend only on the coordinate $z$, the field theory can be formulated
as a functional path integral for a dynamical field $\phi({\bf r},z)$ 
which evolves in a temporal coordinate $z$ \cite{dehocon}. The functional 
Hamiltonians are denoted by $H_E$ in the exterior region, $H_B$ in the 
bulk and $H_S$ in the surface-exclusion layer. In three dimensions this functional problem 
cannot be solved exactly but in one dimension it can be and leads to an
explicit solution for the one-dimensional Coulomb gas \cite{oned}. The
free Debye-H\"uckel theory can be also solved in this formulation \cite{deho}
and one can develop a perturbation theory about it as we shall show here. 
For the moment we will use the Hamiltonian formulation explicitly in order
to find a formal expression for the excess surface tension.
For a globally electro-neutral system with no interfaces and Hamiltonian
$H$ and length $L$ in the $z$ direction, one may write the grand 
partition function as
\begin{equation}
\Xi = {\rm Tr}\exp(-LH)
\end{equation}
that is, we take the system to be periodic in the $z$ direction.
Hence for the pure bulk of electrolyte density $\rho$ one has that for 
large $L$
\begin{equation}
\Xi^{(B)} =   \langle \Psi_0^{(B)}(\rho)| \exp\left(-LH_B(\rho)\right)
|\Psi_0^{(B)}(\rho) \rangle
\end{equation}
and for the exterior region
\begin{equation}
\Xi^{(E)} =   \langle \Psi_0^{(E)}| \exp\left(-L'H_E\right)
|\Psi_0^{(E)} \rangle
\end{equation}
where $ |\Psi_0^{(B)(\rho)} \rangle$ and $|\Psi_0^{(E)} \rangle$ are the 
normalized ground-state wave functionals for the bulk and exterior
functional Hamiltonians $H_B(\rho)$ and $H_E$ respectively. Note that the 
wave functionals must be normalized so that the corresponding grand potential 
is zero for a system of zero volume, that is, zero length in the $z$-direction.   
If the corresponding ground-state energies are  $E_0^{(B)}(\rho)$ and $E_0^{(E)}$,
then we have
\begin{eqnarray}
\beta J^{(B)}(L) &=& L E_0^{(B)}(\rho) \\
\beta J^{(E)}(L') &=& L' E_0^{(E)}
\end{eqnarray}
and the corresponding bulk pressures are given by
\begin{eqnarray}
\beta P^{(B)} &=& -{E_0^{(B)}(\rho)\over A}\label{pbb} \\
\beta P^{(E)} &=& -{ E_0^{(E)}\over A}\label{pbe}
\end{eqnarray}
For the system with interface we find
\begin{equation}
\Xi = \langle \Psi_0^{(E)}|\exp\left(-L'H_E\right)\exp\left(-hH_S\right)
\exp\left(-(L-h)H_B)\right) |\Psi_0^{(B)}(\rho) \rangle
\end{equation}
this is easily seen by joining two such systems together with periodic
boundary conditions. 
We thus obtain
\begin{equation}
\beta J(L,L') = \left[ L' E_0^{(E)} + (L-h)E_0^{(B)}(\rho)
- \ln\left(  \langle \Psi_0^{(E)}|\exp\left(-hH_S\right) 
|\Psi_0^{(B)}(\rho) \rangle \right)\right]
\end{equation}
Using this the excess surface tension is given by
\begin{equation}
\sigma_{e}(\rho) = -{1\over \beta A}\left[ h\left(E_0^{(B)}(\rho)
- E_0^{(B)}(0)\right) + \ln\left({\langle \Psi_0^{(E)}|
\exp\left(-hH_S\right) |\Psi_0^{(B)}(\rho) \rangle
\over \langle \Psi_0^{(E)}|
\exp\left(-hH_S\right) |\Psi_0^{(B)}(0) \rangle}\right)\right]
\end{equation}
Using the relations Eq. (\ref{pbb}) and Eq. (\ref{pbe}) we thus obtain
\begin{equation}
\sigma_{e}(\rho) = h\Delta P(\rho)-{1\over \beta A}\ln\left({\langle \Psi_0^{(E)}|
\exp\left(-hH_S\right) |\Psi_0^{(B)}(\rho) \rangle
\over \langle \Psi_0^{(E)}|
\exp\left(-hH_S\right) |\Psi_0^{(B)}(0) \rangle}\right)
\label{pisig}\end{equation}
where 
\begin{equation}
\Delta P(\rho) = P_B(\rho) - P_B(0)
\end{equation}
is the bulk pressure due to the presence of the electrolyte. The expression
Eq. (\ref{pisig}) is difficult to evaluate, although an approach using standard
quantum-mechanical perturbation theory might be investigated. However,
if the original field theory is free or Gaussian, Eq. (\ref{pisig}) is 
relatively straightforward to compute. We shall use  Eq. (\ref{pisig})
to evaluate the contribution to the surface tension coming from
the cumulant expansion about the free Debye-H\"uckel theory.

\section{Cumulant expansion of the excess surface tension}
Perturbation theory about the Debye-H\"uckel theory \cite{neor} is carried out
by decomposing
the action $S$ in the following manner
\begin{equation}
S = S_0 + \Delta S
\end{equation}
The first term $S_0$ is a Gaussian or free term given by
\begin{eqnarray}
S_0 &=& -{\beta\epsilon_0\over 2} \int_{[-L'-h,-h]\times A}d{\bf x}\ 
(\nabla \phi)^2 -{\beta\epsilon\over 2} \int_{[-h,0]\times A}d{\bf x}\ 
(\nabla \phi)^2 \nonumber \\
  &-&{\beta\epsilon \over 2} \int_{[0,L-h]\times A}d{\bf x}\ 
\ \left((\nabla \phi)^2 + m^2 \phi^2\right) + 2\mu (L-h) A 
\end{eqnarray}
where $m$ is the Debye mass given by $m^2 = 2\rho e^2\beta/\epsilon$. 
The correction to the Gaussian action $\Delta S$ is given by
\begin{equation}
\Delta S = \int_{[0,L-h]\times A} d{\bf x}\ 
\left[ 2 \mu \left (\cos\left(e\beta \phi\right) -1\right)
+ {\beta \epsilon m^2\over 2}\phi^2 \right] 
\end{equation}
The term
$\Delta S$ is  of order of the dimensionless coupling
constant $g = l_B/l_D$ where $l_D = 1/\sqrt{m}$ is the 
Debye length and $l_B = e^2\beta/4\pi\epsilon$ is the Bjerrum length.
A cumulant expansion in $\Delta S$ generates a resummed
expansion in $g$ in the sense that the term of order $n$ in the 
cumulant expansion has the form $C_n = g^n f_n(g)$. In the bulk the function
$f_n(g)$ then has the form $f_n(g) = \sum_{m=1}^\infty a_{n,m} g^m$. However
in the presence of the interface we will see that  $f_n(g)$ has an extra term
containing logarithmic terms in $g$ of type $\sum_{m=1}^\infty a'_{n,m} g^m
\ln(g)$.
This can be shown
by considering the form of the bulk action $S_B$ written in times of
the dimensionless field $\phi' = e\beta/\phi\sqrt{g}$ and by measuring
length in units of the Debye length ($y = mx$). In the new field and length
variables one has the bulk action
\begin{equation}
S_B = -{1\over 2}\int d{\bf y}\ {1\over 4\pi}(\nabla \phi')^2 
+{Z(g)\over 4\pi g}\int d{\bf y} \ \cos(\sqrt{g}\phi')
\end{equation}
where $Z(g)$ is given by $Z(g)= 1/\langle \cos(\sqrt{g}\phi')\rangle$.  Here we
have used the fact that at a point ${\bf x}$ in the system, the average
density of cations/anions is given by
\begin{equation}
\rho_\pm({\bf x}) = 
\mu\langle \exp\left(\pm ie\beta \phi({\bf x})\right) \rangle \label{rhomu}
\end{equation} 
and for ${\bf x}$ in the bulk $\rho_\pm({\bf x}) = \rho$. It is easy to check 
that $Z(g)= 1 + z_1 g + z_2 g^2 \cdots$. Using the same decomposition in the 
bulk as above we obtain
\begin{equation}
S_B = S_0 + \Delta S
\end{equation}
where
\begin{equation}
S_0 = -{1\over 2}\int d{\bf y}\ {1\over 4\pi}\left[(\nabla \phi')^2 + \phi'^2\right] + 
{Z(g)\over 4 \pi g}\int d{\bf y}
\end{equation}
is the Gaussian or free action and
\begin{equation}
\Delta S =  {1\over 4\pi g}\int d{\bf y}\ \left[ Z(g) \cos(\sqrt{g}\phi')
+ {g\over 2}\phi'^2 - Z(g)\right] \label{DS}
\end{equation}
Using the series form for $Z(g)$ we see that $\Delta S$ can be expressed
as a power series in $g$ with first term $O(g)$. It can also be shown that
$\langle \Delta S\rangle = 0$ at $O(g)$ for the homogeneous bulk system;
the corollary is that $\langle \Delta S\rangle \ne 0$ at $O(g)$ only for
systems which are not translationally invariant such as the system with an
interface under discussion here. The outcome is that when calculating to
leading order in $g$ we just need to keep the first term in the cumulant expansion
of the free field theory with $\Delta S$ treated as a perturbation and
the $O(g)$ contributions to $J^{(B)}(L)$ and $J^{(E)}(L')$ in Eq. (\ref{dst})
are zero. We write
\begin{equation}
\Xi = \int d[\phi] \exp\left( S_0 + \Delta S \right) \approx 
\exp\left( \langle \Delta S\rangle _0 \right)
\int d[\phi] \exp\left( S_0 \right)
\end{equation}
with 
\begin{equation}
 \langle \Delta S\rangle _0 = {\int d[\phi] \Delta S\   \exp\left( S_0 \right)
\over \int d[\phi] \exp\left( S_0 \right)}
\end{equation}
This first term in the cumulant expansion can also be shown to begin with the
two-loop term of the standard loop expansion, and hence we shall also refer 
the calculation that follows as the two-loop, or more correctly, the resummed two-loop
calculation.

To this order of approximation the grand potential is given by
\begin{equation}
J = J_0 + \Delta J
\end{equation}
with
\begin{eqnarray}
-\beta J_0 &=& \ln\left( \int d[\phi] \exp\left( S_0 \right)\right)\\
-\beta \Delta J &=& \langle \Delta S\rangle _0 
\end{eqnarray}
The action $S_0$ is Gaussian and we define the correlation 
function of the field $\phi$ at the same point and a distance $z$ from the 
surface-exclusion layer, by
\begin{equation}
\langle \phi({\bf r},z)\phi({\bf r},z)\rangle_0 = G(0,z)
\end{equation}
where we have used the fact that the system is isotropic in 
the plane $A$ (${\bf r} \in A$) but is not isotropic in the
direction $z$. As the action $S_0$ is purely quadratic we also
have that
\begin{equation}
\langle \phi({\bf r},z)\rangle_0 = 0
\end{equation}
For the bulk system (i.e., without an interface) we note that the same-point
field correlator at this level of approximation is given by
\begin{equation}
\langle \phi({\bf r},z)\phi({\bf r},z)\rangle_0 = G_B(0) = G(0,\infty)
\end{equation}
since the physics as $z \to \infty$ for the system with an interface at $z=0$ 
is the same as that of the bulk system.  Using this result, we find that for the 
system with interface
\begin{equation}
\langle \Delta S\rangle _0 = A\int_0^\infty
 dz\ \left[ 2\mu\left(\exp(-{e^2\beta^2
G(0,z)\over2}) -1 \right) + {\beta\epsilon m^2\over 2}G(0,z) \right]
\end{equation}
Using Eq. (\ref{rhomu}) to relate $\rho$ and $\mu$, we find that  to $O(g)$
the fugacity $\mu$ is determined by
\begin{equation}
\rho = \mu \exp\left(-{\beta^2 e^2\over 2} G(0,\infty)\right)
\end{equation}
Using the results above, we find
\begin{equation}
\langle \Delta S\rangle _0 = A\int dz\ \left[ 2\rho\left(\exp(-{e^2\beta^2
G_R(0,z)\over2}) -\exp(-{e^2\beta^2
G(0,\infty)\over2}) \right) + {\beta\epsilon m^2\over 2}G(0,z) \right]
\label{eqds1}
\end{equation}
where we have defined 
\begin{equation}
G_R(0,z) = G(0,z) - G(0,\infty)
\end{equation}
Since we seek a result accurate to $O(g)$ we may expand the second exponential 
in the integral in Eq. ({\ref{eqds1}) to first order and neglect higher  $O(g^2)$ terms. 
Using the definition of the Debye mass $m$ this yields
\begin{equation}
{\langle \Delta S\rangle _0\over A} =  \int dz\ \left[ 2\rho\left(\exp(-{e^2\beta^2
G_R(0,z)\over2}) -1 \right) + {\beta\epsilon m^2\over 2}G_R(0,z) \right]
\label{fdelta}
\end{equation}
The first term in Eq. (\ref{fdelta}) is finite even in the limit $h \to 0$ whereas the
second term
\begin{equation}
\Gamma ={\beta\epsilon m^2\over 2} \int dz G_R(0,z) \label{gam1} 
\end{equation}
is ultra-violet divergent as $h \to 0$. This divergence is due to the integral over
the potential due to the image charge. We might naively resolve this potential difficulty
by observing that if we also expand the first exponential in Eq. (\ref{fdelta}) this 
term is exactly cancelled $\langle \Delta S\rangle _0 =0$ to $O(g)$, so resolving
the difficulty. However, this is expansion is incorrect since this divergence is, in fact, 
cancelled by another arising in $J_0$. The expansion of the first exponential gives
rise to an erroneous divergence which then survives wrongly in the final result; there is
no such divergence. The form of Eq. (\ref{fdelta}) is familiar since the first exponential
is the Boltzmann factor for the repulsive image charge potential that we should expect to
appear and is reminiscent of terms in the the Mayer expansion.

To calculate $G(0,z)$ it is convenient to use the path integral representation
of the problem. Using 
\begin{equation}
\phi({\bf r},z) = {1\over \sqrt{A}}\sum_{\bf p} \tilde{\phi}({\bf p},z)
\exp(i{\bf p}\cdot {\bf r})
\end{equation}
we find that the Gaussian action $S_0$ simply becomes sum of independent
Harmonic oscillators
\begin{equation}
S_0 = 2\mu L +\sum_{{\bf p}} S_{\bf p} 
\end{equation}
where
\begin{equation}
S_{\bf p} = -{1\over 2} \int dz \left[\ M(z) 
{\partial \tilde{\phi}({\bf p})
\over \partial z}{\partial \tilde{\phi}(-{\bf p})
\over \partial z} + M(z) \omega^2({\bf p},z) \tilde{\phi}
({\bf p})\tilde{\phi}(-{\bf p})\right]
\end{equation}
where $M(z) = \beta\epsilon(z)$, $\omega({\bf p},z) = |{\bf p}|= p$
for $z\in [-L', h]$ and $\omega({\bf p},z) = \sqrt{{\bf p}^2 + m^2}$
for $z\in [h, L]$. By expanding in terms of the Fourier modes we find that
\begin{equation}
G(0,z) = {1\over A}\sum_{{\bf p} }\langle \tilde{\phi}({\bf p},z)
\tilde{\phi}(-{\bf p},z)\rangle_0
\end{equation}
The Euclidean Feynman propagator for a simple Harmonic oscillator,
with Hamiltonian denoted by $H_o(\omega,M)$, over a time 
$t$ given by \cite{fehi}
\begin{equation}
\langle X|\exp\left(-tH_o(\omega,M)\right)| Y\rangle =
\left({M\omega\over 2\pi \sinh(\omega t)}\right)^{1\over2}
\exp\left(-{1\over 2}M\omega\coth(\omega t)\left[X^2 + Y^2 - 
2 XY{\rm sech}(\omega t)\right]\right) \label{fpsho}
\end{equation}
and the ground-state wave function is given by
\begin{equation}
\langle \psi_0(\omega,M)|X\rangle  = 
\left( {M\omega\over \pi}\right)^{1\over 4}
\exp\left(-{1\over 2} M\omega X^2\right)\label{gsho}
\end{equation}
with energy $E_0(\omega,M) = \omega/2$. In the free field theory we thus find
\begin{equation}
\langle \tilde{\phi}({\bf p},z)
\tilde{\phi}(-{\bf p},z)\rangle_0 = 
{\langle \psi_0(\omega_E({\bf p}),M_E)| \exp\left(-h H_o(\omega_S({\bf p}),
M_S)\right)
\exp\left(-z H_o(\omega_B({\bf p}),M_B)\right) X^2 |
\psi_0(\omega_B({\bf p}),M_B)\rangle \over
\langle \psi_0(\omega_E({\bf p}),M_E)| \exp\left(-h H_o(\omega_S({\bf p}),
M_S)\right)
\exp\left(-z H_o(\omega_B({\bf p}),M_B)\right)  |
\psi_0(\omega_B({\bf p}),M_B)\rangle}
\end{equation}
where the subscripts $B$, $E$, and $S$ refer to the bulk exterior and
surface-exclusion 
layer values of the various simple harmonic oscillator Hamiltonians
$H_o$ and the corresponding masses $M$ and frequencies $\omega$ 
in these regions.

Carrying out the Gaussian integrations we thus obtain that
\begin{equation}
\langle \tilde{\phi}({\bf p},z) \tilde{\phi}(-{\bf p},z)\rangle_0 = 
D_{33}^{-1}
\end{equation}
where $D$ is the matrix
\begin{equation}
D =  \pmatrix{a & -b & 0\cr
-b & c &- d \cr 0 & -d & e \cr}
\end{equation}
The elements of $D$ are given by
\begin{eqnarray}
a &=& \beta \epsilon_0 p + \beta \epsilon p \ \coth(ph)      \nonumber \\
b &=& \beta\epsilon p \ {\rm cosech}(ph)     \nonumber \\
c &=&  \beta\epsilon p \ \coth(ph) + \beta\epsilon \sqrt{p^2 + m^2}\ \coth(\sqrt{p^2 + m^2}\ z)    \nonumber \\
d &=&   \beta \epsilon  \sqrt{p^2 + m^2} {\rm cosech}(\sqrt{p^2 + m^2}\ z)  
\nonumber \\
e &=&  \beta \epsilon \sqrt{p^2 + m^2}(1+\coth(\sqrt{p^2 + m^2}\ z))
\end{eqnarray}    
A long but straightforward calculation now gives the result
\begin{equation}
G(0,z) = {m\over 2\pi \beta\epsilon}\int dk\ k{K\coth(Kmz) + kB\over
K(kB +K)(1 +\coth(Kmz))} \label{goz}
\end{equation}
where the integral over $k$ is between $0$ and $\Lambda/m$ where $\Lambda$ is 
an ultra-violet cutoff in the Fourier modes of the field $\phi$ in the
plane $A$. In the present calculation we will see there are no ultra-violet 
divergences and we may take the limit $\Lambda\to \infty$. In Eq. (\ref{goz})
and through out the rest of this paper we use the following definitions
\begin{equation}
K = \sqrt{k^2 +1}
\end{equation}
and
\begin{equation}
B = {1 - \Delta\exp(-2kmh)\over 1+\Delta \exp(-2kmh)}
\end{equation}
where 
\begin{equation}
\Delta = {\epsilon-\epsilon_0\over \epsilon_0+\epsilon}
\end{equation}
Using Eq. (\ref{goz}) we find that
\begin{equation}
G(0,\infty) = {m\over 4\pi \beta\epsilon}\int dk\ {k\over K} \label{gozi}
\end{equation}
and using Eq. (\ref{goz}) and Eq. (\ref{gozi}) we obtain
\begin{equation}
G_R(0,z) =  {m\over 4\pi \beta\epsilon}\int dk {k(K-kB)\over K(kB + K)}
\exp(-2Kmz) = {g\over e^2 \beta^2}A(zm) \label{defag}
\end{equation}

In the case $\Delta=1$, Levin and Flores-Mena in Eq. (8) of reference \cite{levfm}
quote a similar formula for $W(z)$ in their notation. Comparing our result at 
$\Delta=1$ with theirs, we note a misprint where the exponential $\exp(-2k(z-d))$ in 
the integrand of their equation should read $\exp(-2pz)$. With this correction we identify
\begin{equation}
W(z) = {g \over 2}A(mz)\Big|_{\Delta=1}\,,~~~\mbox{with}~~~d \equiv h.
\end{equation}
Our result, however, applies for all $\Delta,~~0 \le \Delta \le 1$ and all $h \ge 0$. 

Using Eq. (\ref{defag}) and Eq. (\ref{gam1}) we find
\begin{equation}
\Gamma
= {\rho g \over 2 m }\int dk\ k {K-kB\over K^2(kB + K)}\label{gam2}
\end{equation}

Repeating the above calculation for a pure bulk system, we see that
in the absence of an interface that $G_R(z,0) = 0$ and consequently that the 
corresponding term $\langle \Delta S\rangle_0$ is zero, and so for 
the pure bulk without interface we have to one loop that $J^{(B)}= J_0^{(B)}$.
For a pure exterior system the action is purely Gaussian and $\Delta S^{(E)}=0$
identically, and so to one-loop Eq. (\ref{dst}) becomes
\begin{equation}
\sigma = {1\over A}( \Delta J + J_0 - J_0^{B} - J_0^{E})
\end{equation}
The excess surface tension is thus given by
\begin{equation}
\sigma_{e}(\rho) = \sigma_e^*(\rho) + \sigma_e^{(0)}(\rho)
\end{equation}
where 
\begin{equation}
\sigma_e^*(\rho) = {\Delta J\over A}
\end{equation}
and $\sigma_e^{(0)}(\rho)$ is the excess surface tension for a system with
just the action $S_0$ which can be calculated exactly in the quantum mechanical
formulation as all the simple harmonic oscillators are decoupled. 
We have from Eq. (\ref{pisig})
\begin{eqnarray}
\sigma^{(0)}_e(\rho) &=& 2\mu h -{1\over \beta A}\sum_{\bf p}
\Big[h\left(E_0(\omega_B({\bf p},\rho),M_B) -  
E_0(\omega_B({\bf p},0),M_B)\right)  \nonumber \\
&+& \ln\left(
{\langle \psi_0(\omega_E({\bf p}), M_E)|\exp\left(-h H_o(\omega_S({\bf p}),M_S)
\right)|\psi_0(\omega_B({\bf p},\rho), M_B)\rangle\over 
\langle \psi_0(\omega_E({\bf p}), M_E)|\exp\left(-h H_o(\omega_S({\bf p}),M_S)
\right)|\psi_0(\omega_B({\bf p},0), M_B)\rangle}\right)\Big]
\end{eqnarray}
where we have made explicit the dependence of the  bulk frequencies 
$\omega_B$ on $\rho$, $\omega_B({\bf p};\rho) = \sqrt{p^2 + m^2(\rho)}$. Note 
that the first term in the right-hand side of the above comes from the constant
term in the action $S_0$. 

Using Eqs. (\ref{fpsho}) and (\ref{gsho}) we obtain
\begin{equation}
\sigma_e^{(0)}(\rho) = P_{Debye}h + {\rho g\over m \beta}\int kdk
\left[2\ln\left(1 +{K-k\over 2k}(1+\Delta\exp(-2kmh))\right)
-\ln({K\over k})\right]
\end{equation}
with $P_{Debye}$ the Debye pressure, that is to say the bulk pressure
to $O(g)$, given by
\begin{eqnarray}
\beta P_{Debye} = 2\mu - {1\over 4\pi}\int kdk (K-k) &=& 2\rho - {m^3\over 24\pi}
\nonumber \\
= 2\rho(1-{g\over 6})
\label{pdeb}
\end{eqnarray}
where the rightmost expression in Eq. (\ref{pdeb})  is 
obtained after calculating $\mu$ in terms of $\rho$ \cite{deho}.

Collecting all these contributions we arrive at our final
result for the excess surface tension
\begin{eqnarray}
\beta \sigma_e &=& 2\rho h\left(1 - {g \over 6}\right) +{2\rho\over m}
\int d(mz)\ \left[1- \exp\left(-{g\over 2}A(mz)\right) \right]
\nonumber \\
&+&g{\rho\over 2 m}\int dk\ k\ 
\left[ 4\ln\left(1 +{K-k\over 2k}(1+\Delta\exp(-2kmh))\right)
-2\ln({K\over k})
+ {(kB-K)\over K^2(kB+K)}\right] \label{feq}
\end{eqnarray}
where the function $A(mz)$ as defined by Eq. (\ref{defag}), and we have arranged the
terms to explicitly show the dependence on the dimensionless coupling $g$. we denote the first
term to be the exclusion term, the second to be the depletion term and the third to be the Casimir term.

To evaluate this expression it is convenient to decompose $A(mz)$ into a component
which is singular as $z \to 0$, which gives the direct interaction with the image
charge, and a component finite in this limit:
{\bacol
\bea
A(mz)~=~{\Delta \exp(-2m(z+h))\over 2m(z+h)}&&~+ \nn\\
\int_0^\infty\: d\theta\;\sinh\theta\: \Bigg[&&\left({\exp(-2mz\cosh\theta-2\theta)\,
(1- \Delta^2\exp(-4mh\sinh\theta)) \over 1+\Delta \exp(-2mh\sinh\theta-2\theta)}\right)+\nn\\ 
&&\nn\\
\Delta \exp(-2mz&&\cosh\theta)\left(\exp(-2mh\sinh\theta)-\exp(-2mh\sinh\theta)\right)\Bigg]\;,
\label{az}
\eea
}

where the change of variable $k = \cosh\theta$ has been used.

The result for $\sigma_e$ is correct in perturbation theory to $O(g)$ and holds for
$0 \le \Delta \le 1$ and $h \ge 0$. In the depletion term, the function $A(mz)$ is the 
potential due to the interaction of a charge with its image  and, as is seen above, not only 
includes the screened Coulomb (Yukawa) potential, which is singular as $z \to 0$ when $h=0$, 
but also contains non-singular correction terms which, in particular, are important 
when $h  > 0$. We have derived (\ref{feq}) directly from the perturbation expansion 
for the free energy but the same result would be obtained from the Gibbs adsorption 
isotherm or the G\"untelberg charging process; in both cases a perturbation 
expansion can be obtained for the appropriate quantity which is then appropriately integrated. 
Levin and collaborators \cite{lev} have derived a similar result to Eq. (\ref{feq}) at $O(g)$ for the 
case $\Delta=1, h=0$ but they assume the phenomenological form for $A(mz)$ given by the screened 
Coulomb potential at $h=0$: the first term in Eq. (\ref{az}). As we shall see in the next section, 
the result by Levin \cite{lev} for $\sigma_e$ is numerically similar to ours when evaluated at 
$\Delta=1,h=0$ but for general values of $\Delta,h$ the full result for $A(mz)$ in Eq. (\ref{az}) 
is needed for an accurate calculation of the depletion term. The Casimir term is generated
automatically in the G\"untelberg charging process used by Levin but again to obtain the general
result correct to $O(g)$ presented here, the process must be derived from the perturbation
expansion for the energy density considered as a function of the electric charge $e$. In addition, in
our approach, whatever the method for deriving $\sigma_e$, the perturbation series for $\sigma_e$ can  
systematically calculated to higher orders in $g$ by including terms of higher order using
the cumulant expansion in $\Delta S$, Eq. (\ref{DS}).

In the next two subsections we discuss the consequences of this result.

\subsection{The Onsager-Samaras limiting law}
In this section we shall consider the case where $h=0$ and  the case $\epsilon > \epsilon_0$. 
We show how the Onsager-Samaras limiting law \cite{onsa} for $\sigma_e$ at $\Delta = 1$ 
follows from our result and we derive the generalization to cases where $\Delta < 1$.

When $h=0$ Eq. (\ref{feq}) becomes
\begin{equation}
\beta \sigma_e = 
{2\rho \over m} \int d(u) \left[1-\exp\left(-{g\over 2}A(u)\right)\right] + {\rho g\over 4m}\Delta\;.
\end{equation}

We thus see that the  calculation of $\sigma_e(\rho)$ to the first order in the cumulant 
expansion about the Debye-H\"uckel approximation is divergence free. We now discuss the 
physical origins of these terms. The first term gives a contribution to the surface tension
$\sigma_e^{(D)}$ which can be interpreted as being proportional  to the depletion of solvent 
with respect to the bulk at the interface within the Debye-H\"uckel approximation. This 
term appears in the original Onsager-Samaras calculation where it is  then integrated 
with respect to the fugacity via the Gibbs adsorption equation to obtain the
excess surface tension. In the Debye-H\"uckel approximation it is easy to see that
\begin{equation}
\beta \sigma_e^{(D)} = -\int_0^\infty dz\left(\rho_+(z) + \rho_-(z) - 2\rho \right),
\end{equation}
where $\rho_\pm(z)$ indicates the average cation/anion density at a distance
$z$ from the surface.

When $h=0$ we have
\begin{equation}
A(u) = {\Delta \exp(-2u)\over 2u} + (1- \Delta^2)\int_0^\infty
d\theta \sinh\theta \exp\left(-2u\cosh\theta\right)
\left( { \exp(-2\theta) \over 1+\Delta \exp(-2\theta)}\right).
\label{au}
\end{equation}
We find the asymptotic expansion of $\sigma_{e}$ in the limit of small $g$ to be
\begin{equation}
\beta\sigma_{e} = -{\rho g\Delta\over 2 m}\left[\ln({g\over 2}) + 2 \gamma - 
{3\over 2} -{1\over 2 \Delta^2}(1+\Delta)\left(
2\Delta \ln(2) - (1+\Delta)\ln(1+ \Delta)\right)\right] +O(g^2\ln(g))
\label{lim}
\end{equation} 

When $\Delta = 1$ Eq. (\ref{lim}) is in agreement with the result of 
Onsager and Samaras \cite{onsa}, thus showing that the limiting law is exact up to the
order of the correction indicated in Eq. (\ref{lim}). We note that from
our earlier discussion higher order corrections coming from the cumulant 
expansion will also be $O(g^2)$.   

\subsection{The general case} 
Our results for $\sigma_e$ and $A(mz)$ in Eqs. (\ref{feq}) and (\ref{az}) apply 
generally for all $0 \le \Delta \le 1,~h \ge 0$. When $h$ is non-zero the addition of 
another length scale in the problem renders the derivation of analytical results 
considerably more complicated. The first term of Eq. (\ref{feq}) has a simple 
physical interpretation, it gives a contribution  $ P_{Debye}h$ to $\sigma_e$ which 
can be interpreted as the work done to expel the ions from the surface-exclusion layer into the
bulk. In the limit where $hm \ll 1$ {\em i.e.} $h \ll l_D$ in the second two 
terms of Eq. (\ref{feq}) we can set $h\approx 0$ and recover the 
Eq. (\ref{lim}) for these two terms. We now present some numerical results based on 
the highly accurate Vegas \cite{vegas} integration package.

To carry out the integration over $z$ in Eq. (\ref{feq}) we need to accurately 
determine the function $A(mz)$ given in Eq. (\ref{defag}) and this itself requires
an integration over $k$. The decomposition for $A(mz)$ given in Eq. (\ref{az}) is vital
for good convergence of this latter integral since it is converted to a well behaved
integral over $\theta$ with the singular nature of the potential $A(mz)$ expressed
explicitly. To attempt the integration over $k$ in Eq. (\ref{defag}) numerically 
would not accurately produce this singular behaviour especially in the region where
it is most important, namely as $z \to 0$. We use Vegas to carry out the $d\theta$ integration
in Eq. (\ref{az}) and so accurately determine $A(mz)$ on a discrete set of closely spaced
points for $z$ in the range $0 \le mz \le 4$ and use interpolation to evaluate this function
at intermediate points. To calculate $\sigma_e$ we carry out the separate integrations in 
Eq. (\ref{feq}) again using Vegas. Accurate convergence of the numerical integration in 
all cases is rapid and errors are negligible. In Fig. (\ref{st_0123}) we show the separate
contributions to $\sigma_e$ of the exclusion term, the depletion term and the Casimir term
(respectively the first, second and third terms in Eq. (\ref{feq})) and the total value, 
as a function of solute molarity $0 < x \le 1.0$ for $h=0.0,\,0.1,\,0.2,\,0.3\,\mbox{(nm)}$ and 
a temperature of $20^0\mbox{C}$ and $\Delta = 0.975$, appropriate for water where 
$\epsilon/\epsilon_0 \sim 80$.

From Fig. (\ref{st_0123}) we note that for $0 < h < 0.3\mbox{(nm)}$ the dependence of
$\sigma_e$ on $h$ is rather mild especially at low solute density, for example $0.2$(mol);  
$\sigma_e$ initially decreases with increasing $h$ but then increases as the 
exclusion contribution begins to dominate and the effect of the depletion and Casimir
terms is reduced. Obviously, for larger $h$ the domination of the exclusion term is complete
and $\sigma_e$ with rise linearly with $h$, Eq. (\ref{feq}), for fixed solute density. However,
the range of $h$ considered here is typical of physical films and should be compared with 
the Debye length at solute density of 1(mol) and $T=20^oC$ of $l_D=0.305\mbox{(nm)}$.   

It is interesting to compare the result for the depletion term calculated from the 
full expression for $A(mz)$ given in Eq. (\ref{az}) with that for $A(mz)$ approximated
by the first term: the Yukawa potential. This is relevant because the Yukawa contribution
has an obvious physical significance as the potential for the image charge repulsion and
is the extension to non-zero $h$ of the potential used by Levin (\cite{lev}). We can then
examine the importance of the non-singular correction term in Eq. (\ref{az}), whose origin is
no so phenomenologically obvious. For solute density of 1(mol) we show in Table (\ref{tab1})
the respective contributions of these two calculations as $h$ increases for two values of
$\Delta = 0.975,0.6$ at $T=20^oC$. The first value corresponds to the water-air interface 
with $\epsilon_{H_20}/\epsilon_0 = 80, ~\epsilon_{exterior}/\epsilon_0 = 1$, and the second 
value is for an interface between water and an exterior medium with $\epsilon_{exterior}/\epsilon_0 = 20$.
In each case, the first column gives the contribution when $A(mz)$ is is approximated by the 
Yukawa term and the second column tabulates the contribution when the full expression is 
used for $A(mz)$. For the water/air interface there is negligible difference for $h=0\mbox{(nm)}$ but 
whilst both contributions decrease with $h$ the full result is over five times larger than the 
phenomenological Yukawa approximation suggests. The difference is much more marked
in the case with $\Delta=0.6$ even at $h=0\mbox{(nm)}$ with the full result an order of 
magnitude larger than the Yukawa approximation when $h=0.3\mbox{(nm)}$. Note that the Debye length is
$l_D=0.305\mbox{(nm)}$, comparable with the largest value of $h$ here. These results show that 
in a realistic film, which will generally have a surface layer of thickness in the range discussed
here, the corrections to the Yukawa approximation to the image-charge interaction Eq. (\ref{az})
are overwhelmingly important, especially when the thickness $h \ge l_D$. For higher solute densities
this inequality likely to be easily satisfied. A similar effect occurs for the Casimir term, and 
it is the slower decrease of both these terms with $h$ compared with the phenomenological prediction
that almost exactly balances the increase of the exclusion term with $h$ so that
the $h$ dependence of $\sigma_e$ is relatively weak in the range shown in Fig. (\ref{st_0123})
(for $\Delta = 0.975$ here). An outcome is that the result of Levin \cite{lev} for $\sigma_e$, 
which applies only to the case $h=0\mbox{(nm)}$ $\Delta=1$, is numerically similar to ours for 
$\Delta = 0.975$ but here we have extended the results accurately to general $\Delta$ and $h$ to $O(g)$. 

It should be noted that in the exclusion term in Eq. (\ref{feq}) the Debye-H\"uckel formula
for the pressure has been used and for solute density of 1(mol) and $T=20^oC$
the dimensionless coupling constant is $g=2.33913$, and so the Debye-H\"uckel correction
to the free gas law pressure is nearly 40\%. This indicates that corrections to the bulk 
pressure  at $O(g^2)$ and higher will make a significant contribution at this and 
higher solute densities and, by inference, the higher order corrections to both the
depletion and Casimir terms in Eq. (\ref{feq}) should be calculated. This is the aim of
work in hand.  

In Fig. (\ref{t_dep}) we show the temperature dependence $10^oC \le T \le 30^oC$ for 
different values of $h$ and solute density of 1(mol). It is clear from the $h=0\mbox{(nm)}$
results that the temperature dependence of the depletion and Casimir terms is very weak,
and that the dominant contribution for $h > 0\mbox{(nm)}$ is from the exclusion term and simply
comes from the $T$-dependence of the free gas pressure, $\propto T$, plus the dependence 
of the Debye-H\"uckel correction $\propto T^{-1/2}$.

\bef
\bec
\epsfig{file=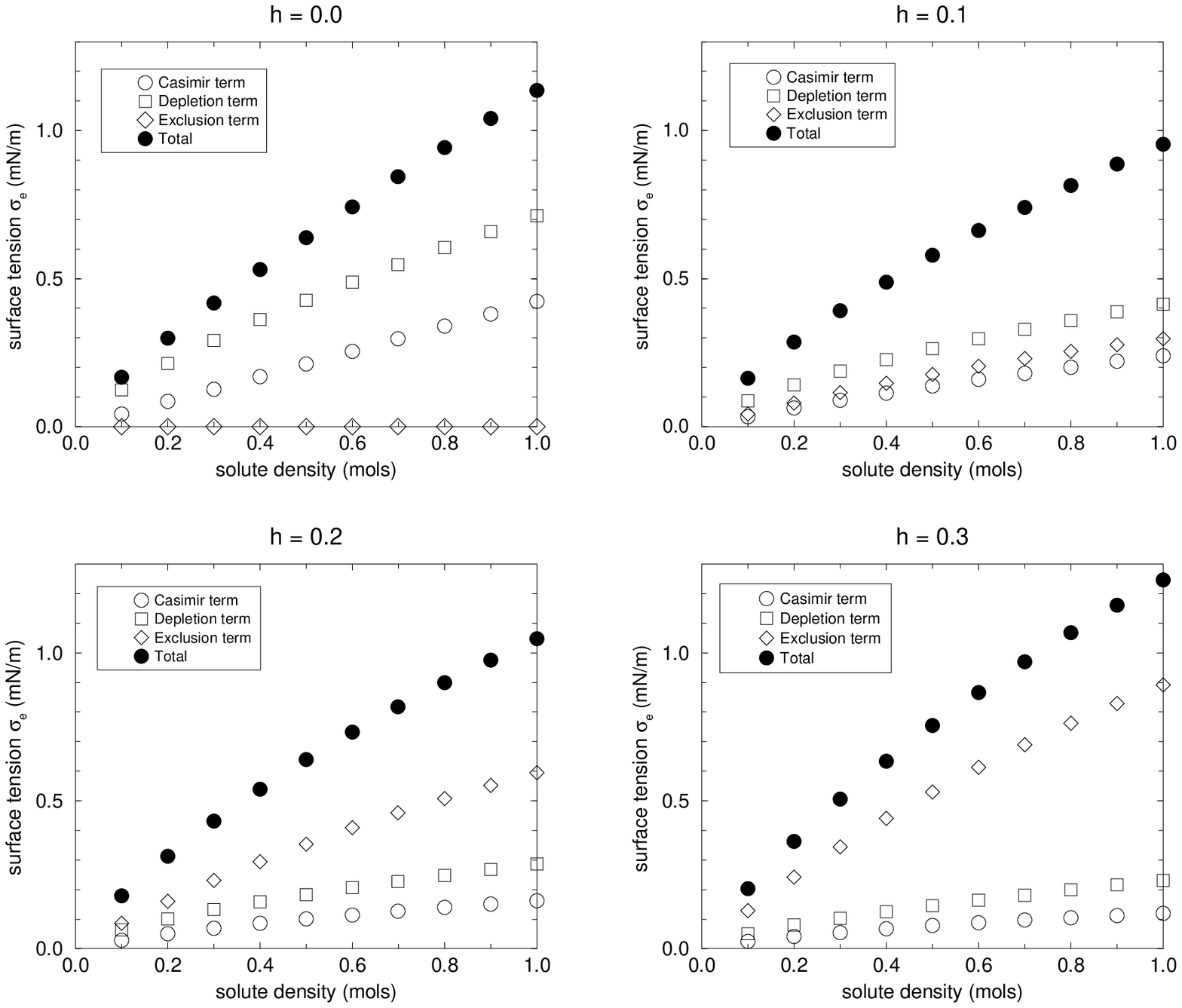,height=120mm}
\enc
\caption{\label{st_0123}\small
The component contributions to the surface tension in mN/m versus the 
molar density for surface exclusion layer thickness $h=0.0,0.1,0.2,0.3\mbox{(nm)}$, $\Delta=0.975$
and $T = 20^oC$.
}
\enf

\btab
\btabu{|c||c|c||c|c|c|}\cline{2-5}
\multicolumn{1}{c}{}&\multicolumn{2}{|c||}{$\Delta = .975$}&\multicolumn{2}{c|}{$\Delta = .6$}\\\hline
$h\mbox{(nm)}$&Yukawa (mN/m)&Full (mN/m)&Yukawa (mN/m)&Full (mN/m)\\\hline
0.0&0.710&0.712&0.561&0.598\\\hline 
0.1&0.286&0.414&0.188&0.331\\\hline 
0.2&0.106&0.287&0.067&0.241\\\hline 
0.3&0.042&0.231&0.026&0.205\\\hline 
\etabu
\caption{\label{tab1}\small
The contribution to $\sigma_e$ of the second (depletion) term in Eq. (\ref{feq}) 
for solute density of 1 (mol) as a function of the exclusion layer thickness
$h$ in (nm). The second column gives the contribution when $A(mz)$ is defined
in Eq. (\ref{az}) is approximated by the first (Yukawa) term and the third column
tabulates the contribution when the full expression is used for $A(mz)$. The results
shown are for two values of $\Delta = 0.975, 0.6$ and $T = 20^oC$. The first value 
corresponds to the water-air interface with $\epsilon_{H_20}/\epsilon_0 = 80,
~\epsilon_{exterior}/\epsilon_0 = 1$, and the second value is for an interface between 
water and an exterior medium with $\epsilon_{exterior}/\epsilon_0 = 20$. For the 
water/air interface there is negligible difference for $h=0\mbox{(nm)}$ but whilst both 
contributions decrease with $h$ the full result is over five times larger than the 
phenomenological Yukawa approximation suggests. The difference is much more marked
in the case with $\Delta=0.6$ with the full result being an order of magnitude larger than
the Yukawa approximation when $h=0.3\mbox{(nm)}$. Note that the Debye length is
$l_D=0.305\mbox{(nm)}$, comparable with the largest value of $h$ here.
}
\etab
\bef
\bec
\epsfig{file=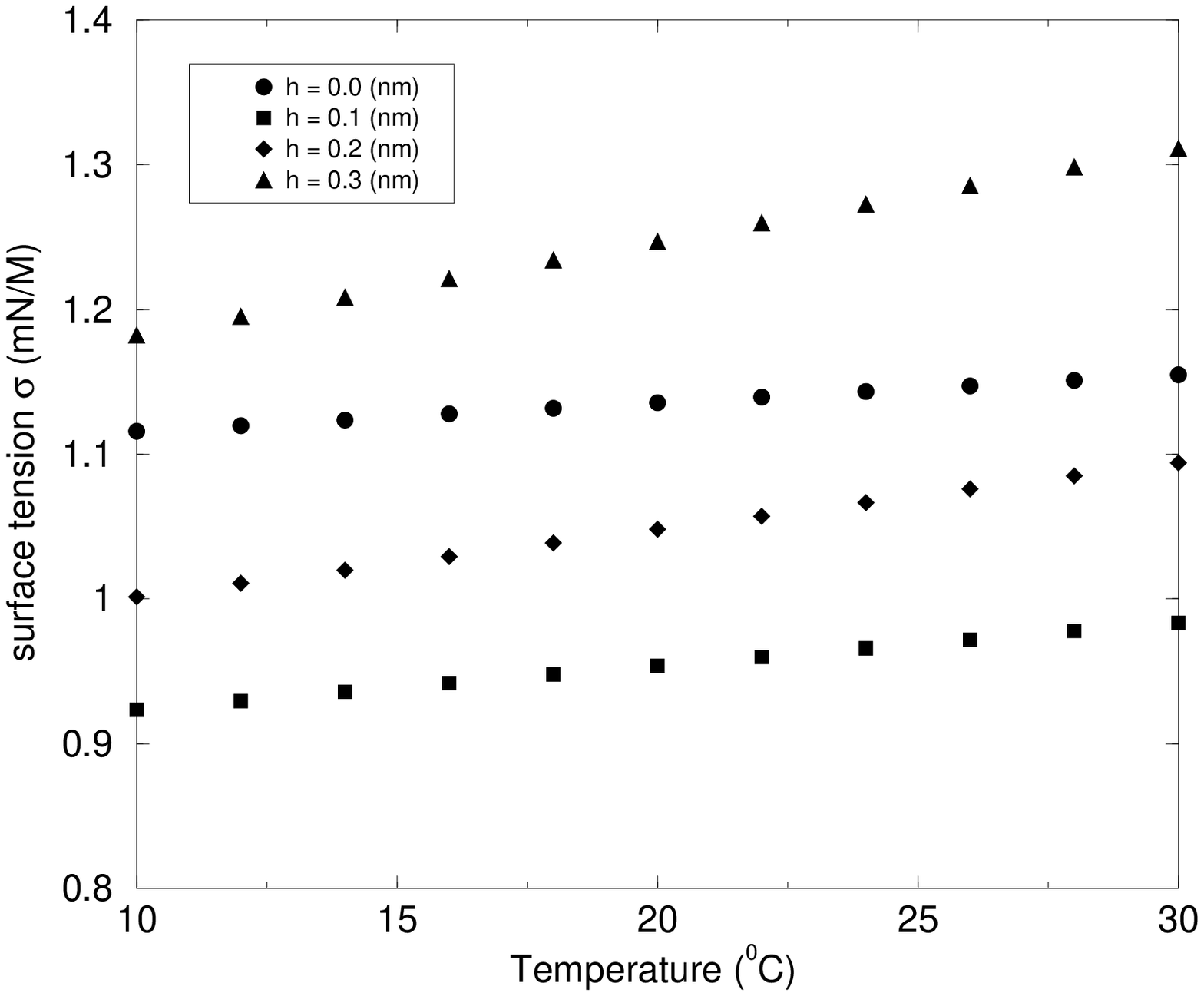,height=60mm}
\enc
\caption{\label{t_dep}\small
The temperature dependence of $\sigma$ for solution of density one mol
for $T$ in range $10-30 (^o)C$ and for surface exclusion layer thicknesses
$h=0.0,0.1,0.2,0.3\mbox{(nm)}$. The dependence on $T$ becomes more marked as
$h$ increases, being about 10\% over this range for $h=0.3\mbox{(nm)}$. This
effect is almost entirely due to the exclusion contribution whose $T$-dependence
comes from the formula for the free gas pressure $\propto T$ plus the  
dependence of the Debye-H\"uckel correction $\propto T^{-1/2}$. Here $\Delta=0.975$.  
}
\enf

%******************************************************************************
% Behavior sigma vs rho for three different values of h, perhaps with OS 
% limiting law shown for the case h=0.
% Behavior of sigma as a function of h at low rho is it monotone ? 
% Behavior of sigma as a function of temperature for three different
% values of h.
% Some comments on the numerical results.
%******************************************************************************

%HERE

\section{Conclusions}
We have shown that the calculation of $\sigma_e$ for a simple model of
an electrolyte can be formulated in terms of a perturbation expansion
in the dimensionless coupling constant $g =l_B/l_D$, where $l_B$ and $l_D$ are
the Bjerrum and Debye lengths, respectively. For the first time we derive 
the full general expressions for $\sigma_e$ to $O(g)$ for general values of 
$\Delta = (\epsilon-\epsilon_0)/(\epsilon+\epsilon_0),~0 < \Delta \le 1$ and
$h \ge 0\mbox{(nm)}$. The calculational method is based on a direct calculation of the 
grand-potential difference between a system with bulk/exterior interface and a bulk 
and exterior system with no interface (both with periodic boundary conditions). In 
this simple model an exclusion layer for the hydrated ions at the surface was 
included, both cations and anions we implicitly taken to be of the same size and 
thus had the same range of exclusion $h$. Due to the symmetry between cations and 
anions in the model here, no mean-field or average electrostatic potential or 
effective surface was generated. The Onsager-Samaras limiting law is shown to be 
the limiting form of the first term in this expansion for small $g$ and for the first time
we have derived its generalization in Eq. (\ref{lim}) to the case where
$\Delta < 1,~h=0\mbox{(nm)}$.

The method is equivalent to other approaches to calculating $\sigma_e$ such as
the Gibbs adsorption isotherm and the G\"untelberg  charging process, both of
which can be formulated using our techniques as perturbation series in $g$. The 
strength of our method is that it gives a systematic expansion which can be 
extended to higher orders in $g$ by including the higher-order terms in the cumulant expansion in 
$\Delta S$ in Eq. (\ref{DS}). The explicit terms given here in Eq. (\ref{feq}) to $O(g)$ are respectively 
the exclusion term due to the exclusion of the gas of solute ions from the surface layer, the 
depletion term which gives the contribution from the image-charge repulsion for ions approaching
the surface, and the Casimir term arising from the change in energy of electric
field modes due to the presence of the surface. Terms higher order in $g$ will correct the 
first two of these contributions. Our method also gives the exact form for these terms,
and especially gives the full expression for the image charge potential $G(z,0) = gA(mz)/e^2\beta$
defined in Eqs. (\ref{defag}) and (\ref{az}). The expected screened Coulomb (Yukawa) potential
used by Levin \cite{lev} can be identified from Eq. (\ref{az}) but the non-singular
correction term is not so easily argued phenomenologically, and from Table (\ref{tab1})
is seen to be important for $h > 0$. By inference a similar effect occurs for the Casimir
term. For the exclusion term in Eq. (\ref{feq}), dominant for large $h$, higher order corrections
to the Debye-H\"uckel approximation are necessary for solute densities greater than 1(mol), and
by inference the other terms will also need correction at higher densities. It should be
noted that Levin \cite{lev} uses the free gas law to compute the exclusion term omitting the
Debye-H\"uckel correction which is equivalent to setting $g=0$ in this term.

The formalism used here can be extended to deal with cases
where an effective surface charge is present due to either a difference 
in hydrated ionic sizes; the behavior of the surface charge in such a model
was recently analyzed in a weak charging approximation by the authors
\cite{dehown}. In addition one may also apply this formalism to other
systems with different energetic or thermodynamic mechanisms leading to 
surface charging; these are the so-called charge regulated models \cite{cr}.
In all cases, one can go beyond the first order expansion used here, although this will
require a more sophisticated theory with a short distance cut-off to 
regularize the divergences arising at higher orders in perturbation theory;
for example, the use of a regularization scheme based on an additional repulsive short range
Yukawa interaction will permit the use of the path integral techniques employed in this 
work.


\newpage
\begin{thebibliography}{0}
\bibitem{expsold}{A. Heydweiller, Ann. Phys. (Leipzig) {\bf 33}, 145 (1910);
Phys. Z. {\bf 3}, 329 (1902)}

\bibitem{expsnew}{P.K. Weissenborn and R.J. Pugh, J. Colloid Interface Sci.,
{\bf 184}, 550 (1996); N. Matsubayasi {\em et al.}, J. Colloid Interface Sci.,
{\bf 209}, 398 (1998)}

\bibitem{wag}{C. Wagner, Phys. Z. {\bf 25}, 474 (1924)}

\bibitem{onsa}{L. Onsager and N.N.T. Samaras, J. Chem. Phys {\bf 2}, 
528 (1934)}

\bibitem{rj}{G. Jones and W.A. Ray, J. Am. Chem. Soc. {\bf 59} 187 (1937);
G. Jones and W.A. Ray, J. Am. Chem. Soc. {\bf 63} 288 (1941);
M. Dole and J.A. Swartout, J. Am. Chem. Soc. {\bf 62} 3039 (1940) } 

\bibitem{dole}{M. Dole, J. Am. Chem. Soc. {\bf 60}, 904 (1939)}

\bibitem{niya}{B.W. Ninham and V. Yaminsky, Langmuir {\bf 13}, 2097 (1997)}

\bibitem{bowini}{M. Bostro\"om, D.R.M. Williams and B.W. Ninham, Langmuir
{\bf 17}, 4475 (2001)}

\bibitem{vlvo}{V.S. Markin and A.G. Volkov, J. Phys. Chem. B {\bf 106}, 11810
(2002)}

\bibitem{kara}{K.A. Karraker and C.J. Radke, Adv. Colloid Interface Sci. 
{\bf 96}, 231 (2002)}

\bibitem{lev}{Y. Levin, J. Stat. Phys. {\bf 110}, 825 (2003); 
Y. Levin, J. Chem. Phys. {\bf 113}, 9722 (2000)}

\bibitem{levfm}{Y. Levin and J.R. Flores-Mena, Europhys. Lett. {\bf 56}, 187 (2001)}

\bibitem{neor}{R.R. Netz and H. Orland, Eur. Phys. J. E {\bf 1}, 203 (2003)}

\bibitem{rand}{J.E.B. Randles, Discuss. Faraday Soc. {\bf 24}, 194 (1957)}

\bibitem{ft1}{R. Podgornik and B. Zeks, J. Chem. Soc. Faraday Trans. II 
{\bf 84}, 611 (1988)}

\bibitem{deho}{D.S. Dean and R.R. Horgan, Phys. Rev. E {\bf 65}, 061603
(2002)} 

\bibitem{dehown}{D.S. Dean and R.R. Horgan,  cond-mat/0210495}

\bibitem{dehocon}{D.S. Dean and R.R. Horgan,  cond-mat/0309028 }

\bibitem{is}{J. Israelachvili, {\em Intermolecular and Surface
forces}, (Academic Press) (1992)}

\bibitem{mani}{J. Mahanty and B. Ninham, {\em Dispersion Forces}
(Academic, London), (1976)}

\bibitem{oned}{S. Edwards and A. Lenard, J. Math. Phys. {\bf 3}, 778,
(1962); D.S. Dean, R.R. Horgan and D. Sentenac, J. Stat. Phys. 
{\bf 90}, 899 (1998)}

\bibitem{fehi}{R. Feynman and A.R.  Hibbs, {\em Quantum Mechanics 
and Path Integrals} (McGraw-Hill) (1965)}

\bibitem{vegas}{G.P. Lepage, {\em Vegas:an adaptive multidimensional integration program}
CLNS-80/447, volume 4, pages 190--195, (1980)}

\bibitem{cr}{B.W. Ninham  and A. Parsegian , J. Theor. Biol 
{\bf 31}, 405, 1971;O. Spalla and L. Belloni, J. Chem. Phys. {\bf 95}, 
7689 (1991); O. Spalla and L. Belloni, Phys. Rev. Lett. {\bf 74}, 2515
(1995); D.S. Dean and D. Sentenac, Europhys. Lett. {\bf 38}, 9, 645, (1997);
D. Sentenac and D.S. Dean, J. Colloid Interface Sci. {\bf 196},
35 (1997);M.N. Tamashoro and P. Pincus, Phys. Rev. E 
{\bf 60}, 6549 (1999); P.A. Pincus and S.A. Safran, Europhys. 
Lett. {\bf 42}, 103 (1998)}



\end{thebibliography}
\end{document}